\documentclass[10pt]{iopart}

\def\p{\partial}
\def\scri{{\mathscr I^+}} 

\usepackage{amssymb,graphicx,color,mathrsfs}

\begin{document}

\title[Exact boundary conditions in numerical relativity \ldots]{Exact
  boundary conditions in numerical relativity using  multiple grids:
  scalar field tests} 

\author{Gioel Calabrese}

\address{School of Mathematics, University of Southampton,
Southampton, SO17 1BJ, UK}

\date{\today}

\begin{abstract}

Cauchy-Characteristic Matching (CCM), the combination of a central
$3+1$ Cauchy code with an exterior characteristic code connected
across a time-like interface, is a promising technique for the
generation and extraction of gravitational waves.  While it provides a
tool for the exact specification of boundary conditions for the Cauchy
evolution, it also allows to follow gravitational radiation all the
way to infinity, where it is unambiguously defined.

We present a new fourth order accurate finite difference CCM scheme
for a first order reduction of the wave equation around a
Schwarzschild black hole in axisymmetry.  The matching at the
interface between the Cauchy and the characteristic regions is done by
transfering appropriate characteristic/null variables.  Numerical
experiments indicate that the algorithm is fourth order convergent.  As
an application we reproduce the expected late-time tail decay for the
scalar field.

\end{abstract}



\section{Introduction}

In the past year there has been remarkable progress in the field of
numerical relativity, particularly in the simulation of binary black
hole systems.  Some of today's codes are capable of tracking the holes
for many orbits before the merger and ringdown
\cite{P,Goddard,Brownsville}.  Since a primary goal of numerical
relativity is the production of gravitational wave templates
associated with the merger of black holes, it is important to
simultaneously make progress with the tools which allow for their
accurate generation and extraction.

The schemes which have been able to push forward the binary black hole
problem either use spatial compactification \cite{P}, which brings
spatial infinity at a finite coordinate location, or mesh refinement
\cite{Goddard,Brownsville}, which allows the boundary to be placed
further out by using lower resolution far away from the strong field
region.  A limitation of both methods is that the number of
grid-points per wavelength gradually diminishes as the radiation
propagates away from the sources and therefore the quality of the
gravitational radiation progressively deteriorates, forcing the
extraction to occur at a relatively small distance from the final
hole\footnote{Another reason for extracting radiation at small
distances is to avoid boundary effects.  Nevertheless, the comparison
of the real part of $r\psi_4$ extracted at $r=20M$ and $r=40M$ done by
Baker {\it et al.}  \cite{Goddard} seems to indicate that the
extraction at small distances leads to sufficiently accurate
results.}.  The undesirable effect of a time-like boundary with a
chronological future that intersects the region of interest (where
physical quantities are calculated), was highlighted in \cite{Lior,DR}.

\begin{figure}[ht]
\begin{center}
\includegraphics*[width=10cm]{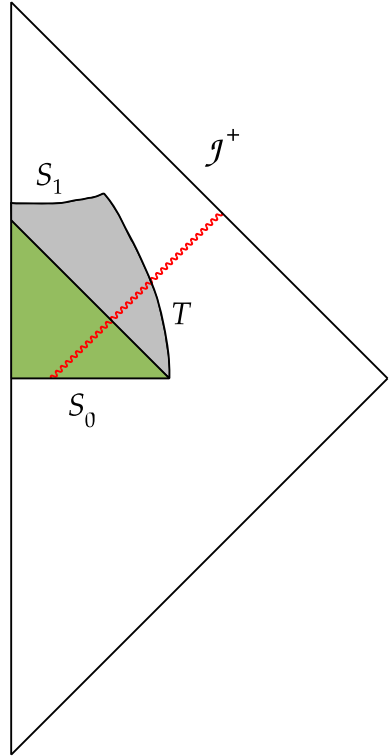}
\caption{Most simulations in numerical relativity are carried out on a
 domain bounded by two space-like slices, $S_0$ and $S_1$ (the initial
 and final time), and a time-like artificial boundary $T$.  Since it is not
 known how to specify local boundary conditions which do not introduce
 spurious radiation or reflect outgoing waves, the accuracy of the
 solution is limited to the future domain of dependence of $S_0$
 (green region).  The solution in the grey region, which is causally
 connected to the boundary, is inaccurate.  In this situation one is
 forced to extract gravitational radiation relatively close to the
 sources.}
\label{Fig:conformal1}
\end{center}
\end{figure}

\begin{figure}[ht]
\begin{center}
\includegraphics*[width=10cm]{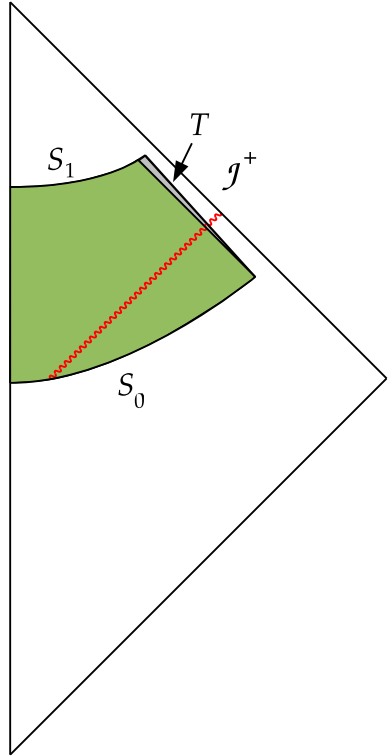}
\caption{When asymptotically null slices are used, as in \cite{CGH},
  the outer boundary $T$ can be placed far away from the sources and
  is approximately null (the incoming modes travel very slowly).
  Gravitational radiation can be accurately extracted at much larger
  distances and with lower computational cost than with standard
  space-like slices. See Table I of \cite{CGH} for a comparison of
  computational costs and errors with different slices.}
\label{Fig:conformal2}
\end{center}
\end{figure}

\begin{figure}[ht]
\begin{center}
\includegraphics*[width=10cm]{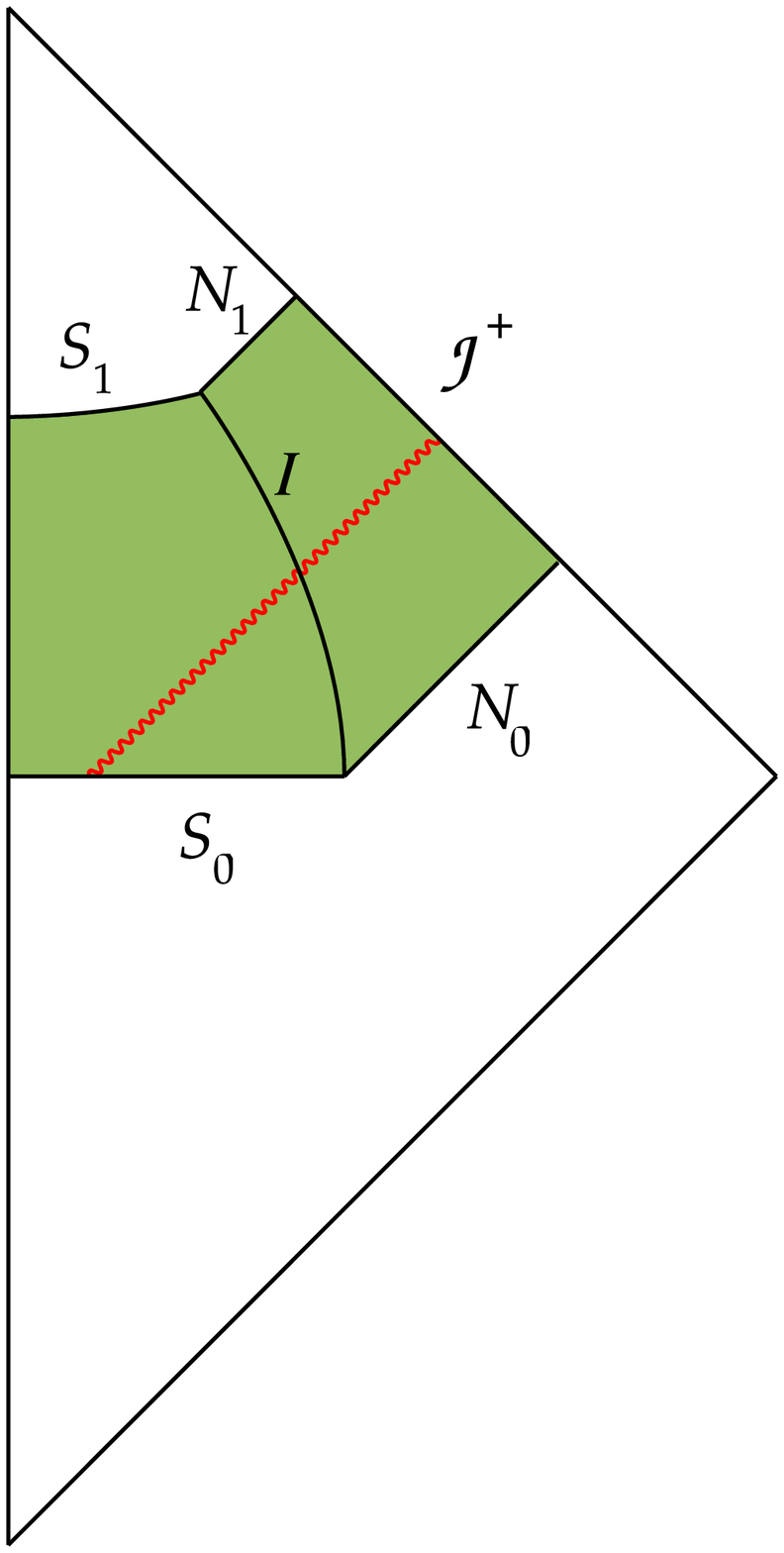}
\caption{By combining a Cauchy and a characteristic scheme one solves
for the entire spacetime of interest, including $\scri$ (future null
infinity), where the radiation is unambiguously defined. Initial data
is specified at $S_0$ and $N_0$ and the two schemes are connected by a
time-like interface $I$.  Although in principle one could use the
asymptotically null approach up to null infinity, experiments with the
wave equation in spherical symmetry suggest that the scheme is less
effective than that which uses a time-like boundary at very large
distances \cite{CGH}.}
\label{Fig:conformal3}
\end{center}
\end{figure}

To improve the quality of the outgoing radiation one should aim to
keep the coordinate distance that the waves have to travel to a
minimum, without making their speed go to zero.  This can be achieved,
for example, by employing asymptotically null slices
\cite{Fra,Kan,M1,M2,M3}, or by supplementing the Cauchy evolution with
a characteristic code
\cite{Sa,BBM,W,dI,CdI,CdIV,DdIC,dIV3,dIV4,DdIV,dIDS,SSV1,SSV2,BGHMPW1,BGHMPW2,GLPW},
which extends all the way to infinity.  See Figures
\ref{Fig:conformal1}, \ref{Fig:conformal2}, and \ref{Fig:conformal3}.
Both methods resolve incoming waves poorly at very large distances,
but these are not expected to be significant.  The reason for not
using a single characteristic evolution scheme, without a Cauchy
interior covering the strong field region, is that such a scheme tends
to develop caustics, where neighbouring characteristics focus and the
coordinate system breaks down.

In this work we describe a fourth order accurate algorithm for the
axisymmetric massless Klein-Gordon scalar field around a Schwarzschild
black hole, which employs a Cauchy evolution scheme, based on
space-like slices, and a characteristic scheme, based on null slices.
The outer boundary of the interior Cauchy region is a time-like
surface and coincides with the inner boundary of the characteristic
region.  Communication between the two regions is done taking into
account the propagation of the characteristic variables.  At the
discrete level the interface is implemented with touching grids.  We
find that when the slices are not simultaneous, this particular
multi-grid approach is better suited than overlapping grids, as it
avoids complications due to causality.  The radial coordinate of the
characteristic region is compactified in order to reach future null
infinity.

The paper is organised as follows.  In Section \ref{Sec:continuum} we
describe the continuum scalar field initial-boundary value problem in
the Cauchy and characteristic regions and discuss the interface
treatment.  In Section \ref{Sec:discrete} we provide details of the
discretization method and in Section \ref{Sec:tests} we describe the
tests carried out to validate the code, including the simulation of
the late-time tail decay.

\section{Klein-Gordon scalar field around a Schwarzschild black hole}
\label{Sec:continuum}

In this section we describe the continuum problem, giving the explicit
form of the equations, and the interface treatment.
The Schwarzschild line element in Kerr-Schild coordinates takes the form
\begin{equation}
ds^2 = -dt^2 + dr^2 + \frac{2M}{r} (dt+dr)^2 + r^2 d\Omega^2, \label{Eq:KS}
\end{equation}
where $M$ is the mass of the black hole and $d\Omega^2 = d\theta^2 +
\sin\theta^2 d\phi^2$. 
The surface defined by
\begin{equation}
t = r + 4M\ln(r-2M) + {\rm const}
\end{equation}
is null and its normal, provided it belongs to the future null cone,
points toward increasing values of $r$.  It is convenient to introduce
the shorthand $r^{\pm} = r\pm 2M$.  Using $\Psi = r\Phi$, in the
Cauchy region the wave equation on the Schwarzschild background,
$\nabla^{\mu}\nabla_{\mu} \Phi = 0$, takes the form
\begin{equation}
\fl\p_t^2 \Psi = \frac{4M}{r^+} \p_r\p_t \Psi + \frac{r^-}{r^+} \p_r^2
\Psi - \frac{2M}{rr^+} \p_t \Psi +\frac{2M}{rr^+} \p_r \Psi -
\frac{2M}{r^+r^2}\Psi + \frac{\p_{\theta}(\sin\theta
  \p_{\theta}\Psi)}{rr^+\sin\theta}. \label{Eq:waveCauchy}
\end{equation}

In Bondi coordinates the Schwarzschild line element is
\begin{equation}
ds^2 = \left(-1+\frac{2M}{r}\right) du^2 - 2\, du \, dr + r^2
d\Omega^2  \label{Eq:Bondi}
\end{equation}
and it is related to (\ref{Eq:KS}) via the coordinate transformation
\begin{eqnarray}
u &=& t-r-4M\ln(r-2M),\\
\tilde r &=& r,\\
\tilde \theta &=& \theta,\\
\tilde \phi &=& \phi,
\end{eqnarray}
which implies
\begin{eqnarray}
\p_t &=& \p_u,\\
\p_r &=& -\frac{\tilde r+2M}{\tilde r - 2M} \p_u + \p_{\tilde r},\\
\p_\theta &=& \p_{\tilde \theta},\\
\p_\phi &=& \p_{\tilde \phi}.
\end{eqnarray}
Where there is no cause for confusion we will not distinguish
between $\{\tilde r,\tilde \theta,\tilde\phi\}$ and
$\{r,\theta,\phi\}$.

The Klein-Gordon equation in the characteristic region becomes
\begin{equation}
0 = -\frac{2r}{r^+} \p_u\p_{\tilde r} \Psi +
\frac{r^-}{r^+}\p_{\tilde r}^2 \Psi + \frac{2M}{rr^+} \p_{\tilde r}
\Psi - \frac{2M}{r^+r^2}\Psi + \frac{\p_{\tilde\theta}(\sin\theta \p_{\tilde\theta}\Psi) }{rr^+\sin\theta}.
\label{Eq:wavechar}
\end{equation}
Note the lack of second partial derivatives $\p_u^2$, reflecting
the fact that the $u={\rm const}$ surface is characteristic.

\subsection{Axisymmetry and first order reductions}

We focus on axisymmetric solutions of the equations
(\ref{Eq:waveCauchy}) and (\ref{Eq:wavechar}), as we do not expect
additional difficulties to arise in the general 3-dimensional case.
This implies $\p_{\phi} \Phi = 0$ and regularity demands that, across
the axis $\theta = 0$ and $\theta = \pi$, the scalar field is an even
function of $\theta$.

Since the discretization of second order systems tends to be more
subtle than that of fully first order systems
\cite{SKW,MBSKW,C1,CHH,CG}, we perform a reduction to first order and
discretise the resulting system.  In the Cauchy region we introduce
the auxiliary variables $T= \p_t \Psi$, $R= \p_r \Psi$ and $\Theta =
\p_{\theta} \Psi$.  The reduced system is
\begin{eqnarray}
\p_t T &=& \frac{4M}{r^+}\p_r T + \frac{r^-}{r^+} \p_r
R - \frac{2M}{rr^+} T +\frac{2M}{rr^+} R - \frac{2M}{r^2r^+}\Psi +
\frac{\p_{\theta}(\sin\theta \Theta)}{rr^+\sin\theta},\\ 
\p_t R &=& \p_r T,\\
\p_t \Theta &=& \p_\theta T,\\
\p_t \Psi &=& T.
\end{eqnarray}
We also have the constraint ${\cal C} \equiv \p_r \Theta - \p_{\theta}
R = 0$, which propagates with zero speed and therefore does not
introduce complications at the boundary.  This system is symmetrizable
hyperbolic provided that $r>2M$ \cite{CLNPRST}.  It has non trivial
characteristic speeds $1$ and $-r^-/r^+$ in the radial direction and
the corresponding characteristic variables are
\begin{equation}
w^{\pm} = \frac{r^+}{r^-} T \pm R.
\end{equation}

To reduce equation (\ref{Eq:wavechar}) to first order we introduce
$\tilde R = \p_{\tilde r} \Psi$, $V = 2r/r^- \p_u \Psi - \p_{\tilde r}
\Psi$ and $\tilde\Theta = \p_{\tilde\theta}\Psi$.  The reason for this
choice of auxiliary variables is that $\p_{\tilde r}$ and $2r/r^-\p_u
- \p_{\tilde r}$ are null vectors.  We will refer to $\tilde R$ and
$V$ as null variables.  The reduced system is
\begin{eqnarray}
\p_{\tilde r} V &=& -\frac{2M}{rr^-} V - \frac{2M}{r^2r^-} \Psi +
\frac{\p_{\tilde \theta}(\sin\theta\tilde\Theta)}{rr^-\sin\theta}, \label{Eq:reduced1}\\
2\p_u \tilde R &=& \frac{r^-}{r} \p_{\tilde r} \tilde R +
\frac{2M}{r^2} \tilde R - \frac{2M}{r^3} \Psi +
\frac{\p_{\tilde\theta} (\sin \theta \tilde\Theta)}{r^2\sin\theta},\label{Eq:reduced2}\\
2\p_u \tilde\Theta &=& \frac{r^-}{r} (\p_{\tilde \theta} V\label{Eq:reduced3}
+\p_{\tilde\theta} R),\\
2\p_u \Psi &=& \frac{r^-}{r} (V+\tilde R).\label{Eq:reduced4}
\end{eqnarray}

To reach $\scri$ we compactify the characteristic radial
coordinate $\tilde r$ by introducing a new independent variable $\xi$
such that
\begin{eqnarray}
\tilde r &=& r_i + \frac{2}{\pi}(\xi_{\infty}-\xi_i) \tan
\left(\frac{\pi}{2} \frac{\xi - \xi_i}{\xi_{\infty}-\xi_i}\right).
\end{eqnarray}
In the last equation $r=r_i$ and $\xi=\xi_i$ represent the location of
the interface boundary in the Cauchy and characteristic coordinate
systems, respectively.  The surface $\xi = \xi_{\infty}$ corresponds to
future null infinity.  Effectively, this amounts to replacing in system
(\ref{Eq:reduced1})--(\ref{Eq:reduced4}) the partial derivatives with
respect to $\tilde r$ with
\begin{equation}
\p_{\tilde r} = \frac{\p \xi}{\p \tilde r} \p_\xi = \cos^2
\left(\frac{\pi}{2} \frac{\xi - \xi_i}{\xi_{\infty}-\xi_i}\right) \p_\xi
\end{equation}
There are coefficients in equation (\ref{Eq:reduced1}) that contain
indetermined forms, which can be explicitly evaluated using the limit
\begin{equation}
\lim_{\xi\to \xi_{\infty}} \tilde r^2 \frac{\p\xi}{\p\tilde r} =
\frac{4}{\pi^2} (\xi_{\infty}-\xi_i)^2. \label{Eq:indetermined}
\end{equation}
We typically use $r_i = \xi_i = 10M$ and $\xi_{\infty} = 20M$.

\subsection{Interface treatment}

A necessary condition for well-posedness for the interior Cauchy
problem is that at the boundary and at the interface data is
prescribed to the incoming characteristic variables.  If no data is
specified to these variables, then one can expect to lose uniqueness.  The
characteristic region can provide accurate data for the incoming mode
by setting
\begin{equation}
w^+|_{r=r_i} = \tilde R|_{\xi=\xi_i}.\label{Eq:wp}
\end{equation}
On the other hand, the characteristic problem requires data for the
outgoing null variable $V$.  This is done by setting 
\begin{equation}
V|_{\xi=\xi_i} = [T - R]_{r=r_i}.\label{Eq:V}
\end{equation}

\subsection{Treatment of future null infinity}
Future null infinity corresponds to the boundary surface $\xi =
\xi_{\infty}$.  Here, as a result of the compactification, the speed
of the incoming variable $\tilde R$ approaches zero, so there are no
incoming modes (the surface is null).  No data is required at
this boundary and its numerical treatment is particularly simple.
See subsection \ref{Sec:scri}.

\section{Finite difference scheme}
\label{Sec:discrete}

Stable finite difference approximations of the initial-boundary value
problem for the wave equation in the Cauchy region (not coupled to the
characteristic problem) are available.  For example, by approximating
the spatial derivatives using finite difference operators that satisfy
the summation by parts rule \cite{St}, it is possible to construct
schemes which conserve a discrete energy \cite{CLNPRST,CLRST}, a
positive definite quadratic form of the dependent variables.  Boundary
conditions can then be imposed either using Olsson's method \cite{Ols1,Ols2,Ols3}
or using penalty terms \cite{CGA,CNG,NC}.

It is possible to obtain energy estimates
in the characteristic region \cite{F1,F2,F3}, and in turn for the
coupled Cauchy-characteristic problem.  Ideally, one would like to
achieve similar estimates at the discrete level.  Although we have
been able to formally extend these estimates to the discrete case, the
semi-discrete problem that we consider fails to admit solutions for
generic initial data.  

Formally, one can obtain discrete estimates for the semi-discrete
system
\begin{equation}
A^u \p_u w + A^x D_x w + A^yD_y w + A^zD_zw= 0, \label{Eq:semidiscrete}
\end{equation}
where $w$, which is a shorthand for $w_{ijk}(u)$, is a vector-valued
grid-function having the dynamical variables as components and the
variables $x$, $y$ and $z$ are discretised with indices $i$, $j$ and
$k$.  Furthermore, $u$ is left continuous, $y$ and $z$ are periodic,
the matrices are constant and symmetric and $A^u$ is singular.  The
operators $D_x$, $D_y$ and $D_z$ are discrete finite difference
operators.  The operators $D_y$ and $D_z$ are centered finite
difference approximations of $\p_y$ and $\p_z$.  The operator $D_x$
approximates $\p_x$ and satisfies the summation by parts
rule\footnote{The summation by parts rule is the discrete analogue of
the integration by parts rule: $\int_a^b w'(x) v(x) + w(x)v'(x) \, dx
= w(b)v(b) - w(a)v(a)$.}
\[
\sum_{i=0}^N (D_xw_i)^T v_i \sigma_i h + 
\sum_{i=0}^N w_i (D_x v_i) \sigma_i h =
w_Nv_N-w_0v_0,
\]
where $\sigma_i$ are appropriate weights (explicit examples of weights
and operators can be found in \cite{St}) and $h$ is the spatial mesh
spacing.  System (\ref{Eq:semidiscrete}) represents the semi-discrete
approximation of (\ref{Eq:reduced1})--(\ref{Eq:reduced4}).  Without
loss of generality we can assume that $A^u$ is diagonal and therefore
at least one equation in system (\ref{Eq:semidiscrete}) does not
contain derivatives with respect to $u$.  Multiplying on the left by
$w^T$ and using the fact that the matrices are constant and symmetric,
we get
\begin{eqnarray}
&&(\p_u w)^T A^u w + w^TA^u \p_u w + (D_xw)^TA^x w + w^T A^x D_x w \nonumber\\
&& + (D_yw)^TA^y w + w^TA^yD_y w + (D_zw)^TA^z w + w^TA^zD_z w = 0.
\end{eqnarray}
Summing over the indices corresponding to $x$, $y$ and $z$ and integrating
in $u$ gives the equality
\begin{equation}
\left[\sum_{i=0}^N w_i^T A^u w_i \sigma_i h\right]_{u=u_0}^{u=u_1} + \int_{u=u_0}^{u=u_1} \left[w_i^T A^x
w_i \right]_{i=0}^{i=N} du = 0,
\end{equation}
where $[f(u)]_{u=u_0}^{u=u_1} = f(u_1)-f(u_0)$, $[f_i]_{i=0}^{i=N} =
f_N-f_0$, the index $i$ corresponds to the $x$ direction, and the
indices corresponding to $y$ and $z$ are suppressed.  Despite the
estimate, a solution of (\ref{Eq:semidiscrete}) does not exist.  This
can be seen by looking at the equation that does not involve
derivatives in the $u$ direction.  The system $D_x w = F$, for generic
$F$, is overdetermined because the operator $D_x$ is singular.

Since the summation by parts approach does not appear to work, our
guiding principle will be to discretise the problem in a way
consistent with the continuum problem\footnote{A finite difference
scheme is said to be consistent if solutions to the continuum problem
satisfy the scheme to an order greater or equal to 1 in the mesh
spacing.}, for which estimates do exist.  Numerical experiments will
be crucial for establishing convergence of the resulting scheme.

\subsection{Cauchy grid}

The discretization of the Cauchy problem is done in a standard way:
the derivatives are approximated with centered fourth order accurate
finite difference operators
\begin{equation}
D_0 \left(1 - \frac{h^2}{6}D_+D_-\right), \label{Eq:fourth}
\end{equation}
where $D_0 w_i = (w_{i+1}-w_{i-1})/(2h)$, $D_+ w_i = (w_{i+1}-w_i)/h$,
and $D_-w_i = (w_i-w_{i-1})/h$.  When applied to a grid-function
$w_i$, operator (\ref{Eq:fourth}) gives
\[
(-w_{i+2}+8w_{i+1}-8w_{i-1}+w_{i-2})/(12h).
\]

The boundaries, the interface and the axis of symmetry are implemented
using ghost points.  To eliminate the black hole singularity from the
domain we use excision.  At the boundary the ghost points are
populated\footnote{Populating ghost points means assigning the values
of the dynamical variables at these points.} using extrapolation and
by overwriting the incoming characteristic fields, if any are present,
at the boundary point.  See Appendix B.2 of \cite{CG} and \cite{GKO}.
Near the axis of symmetry these are populated using the regularity
conditions, i.e., using the fact that $\Psi$ is an even function of
$\theta$ across the axis.  The interface boundary is treated as if it
was an outer boundary, with the only exception that the incoming mode
is provided by the characteristic evolution.

We also add some artificial dissipation to the scheme of the form
$\sigma h^5 (D_+D_-)^3$, explicitly
\[
\sigma(w_{i+3}-6 w_{i+2}+15 w_{i+1}-20 w_{i}+15 w_{i-1}-6 w_{i-2}+w_{i-3})/h, 
\]
with $\sigma = 0.075$ \cite{CN2} and use fourth order Runge-Kutta to
integrate in time with a Courant factor of $0.5$.  Although we are
investigating a linear problem, the partial differential equations
have variable coefficients.  A certain amount of artificial
dissipation is desirable for stability and will most likely be
necessary in the non-linear case.

\subsection{Characteristic grid}

The discretization of the evolution part of the system,
Eqs.~(\ref{Eq:reduced2})--(\ref{Eq:reduced4}), is done as in the
Cauchy case.  Equation (\ref{Eq:reduced1}) treated as an ordinary
differential equation in $\tilde r$, the last two terms in the right
hand side taking the role of forcing terms.  It is advanced in the
$\tilde r$ direction by using fourth order Runge-Kutta (with a
``time'' step equal to the radial mesh spacing).  This requires the
evaluation of the coefficients and forcing terms of
(\ref{Eq:reduced1}) at intermediate time steps.  The missing data is
generated with fourth order interpolation
\begin{equation}
U_{i+1/2} = (-U_{i-1}+9U_i+9U_{i+1}-U_{i+2})/16.
\end{equation}
Note that to start the integration we need the fields at the grid-point
$i=-1$.  We use fourth order extrapolation to populate this point.

The integration of $V$ along the $\tilde r$ direction is done at each
intermediate time step of the global time integrator.  Artificial
dissipation is added to the evolution in the $u$ direction.

\subsection{Axis of symmetry}

We use ghost points to treat the axis $\theta = 0,\pi$.  The dynamical
variables $\Psi$, $T$, $R$, $\tilde R$ and $V$ are even functions of
$\theta$ along the axis, and $\Theta$ and $\tilde \Theta$ are odd
functions.  In particular, we impose that $\Theta = \tilde \Theta = 0$
on the axis.

\subsection{Interface treatment}

Data is communicated at the interface at each intermediate time step:
the Cauchy code requires the incoming variable $w^+$,
Eq.~(\ref{Eq:wp}), and the characteristic code needs the outgoing null
variable $V$, Eq.~(\ref{Eq:V}).  The Cauchy and characteristic grids
are touching, rather than overlapping.  This avoids complications with
causality which would arise if the $u={\rm const.}$ slice intersected
the domain of dependence of the $t={\rm const}$ slice.

At the interface boundary we implement condition (\ref{Eq:wp}) by
solving for $T$ and $R$ the system of equations
\begin{eqnarray}
\frac{r^+}{r^-} T + R &=& \tilde R,\\
\frac{r^+}{r^-} T - R &=& \frac{r^+}{r^-} T_{\rm old} - R_{\rm old}.
\end{eqnarray}

\subsection{Treatment of future null infinity}
\label{Sec:scri}

The only difficulty at the outer boundary of the characteristic grid
is that some coefficients contain indetermined forms ``$0 \cdot
\infty$''.  We use the limit (\ref{Eq:indetermined}) to evaluate
these.  We populate the ghost points by extrapolating all fields.  As
there are no incoming modes no further action is taken.

\subsection{The algorithm}

In this subsection we list the basic steps of our algorithm.
\begin{enumerate}
\item Give initial data to all fields in the Cauchy grid ($\Psi$ and
  $T$ can be given arbitrarily, but $R$ and $\Theta$ have to obey
  their definition constraints).  Give initial data to all fields on
  the characteristic grid ($\Psi$ can be given arbitrarily, $\tilde R$
  and $\tilde \Theta$ have to satisfy their definition constraints), with
  the exception of the null variable $V$.
\item Impose regularity on the axis, impose interface boundary
  conditions (\ref{Eq:wp}) and (\ref{Eq:V}).  Integrate
  Eq.~(\ref{Eq:reduced1}) using fourth order Runge-Kutta (use fourth
  order Lagrange interpolation to populate the coefficients and
  forcing terms of the differential equation). Extrapolate all fields
  in the radial direction to populate the ghost points.  Impose
  regularity again.\label{item:2}
\item Evaluate the right hand sides of the evolution equations
  (because we use ghost points this can be done at every interior, and
  boundary/axis/interface point) and calculate the solution at the
  next intermediate time step.
\item Repeat from point (\ref{item:2}) until a full fourth order
  Runge-Kutta time step is taken.
\item Repeat from point (\ref{item:2}) and take as many Runge-Kutta time
  steps are needed to reach the final time.
\end{enumerate}

\section{Numerical tests}
\label{Sec:tests}

\subsection{Convergence test}

To test for convergence we set $M=0$, construct an exact solution
by translating the spherically symmetric solution $\Phi = f(t-r)/r$
along the axis by an amount $\delta$, 
\begin{equation}
\Psi = r \Phi = \frac{r}{\hat r} f(t-\hat r),
\end{equation}
where $\hat r^2 = r^2 - 2r\delta\cos\theta + \delta^2$. The
expressions for $T$, $R$, $\Theta$ and $\tilde R$, $V$, $\tilde
\Theta$ can be readily computed.  We monitor how the $L_2$-norm of the
error between the numerical and exact solution,
\[
\| w-u_{({\rm exact})} \| = \left( \sum_j (w_j-u_{({\rm exact})})^2 h_r h_{\theta}\right)^{1/2},
\]
scales with resolution.  Since the exact solution is known, we only
need to use two resolutions (coarse and fine) to test for convergence.
In the expression above $h_r$ and $h_{\theta}$ represent the mesh
spacing and the sum is extended to all grid-points of the Cauchy and
charateristic grids.

\begin{figure}[ht]
\begin{center}
\includegraphics*[width=12cm]{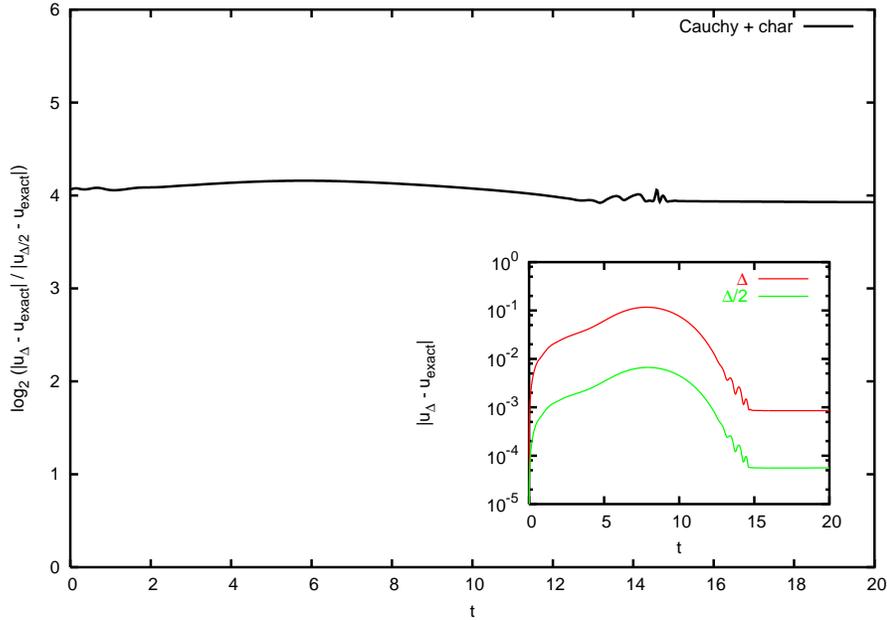}
\caption{This plot illustrates the experimental convergence rate of
  the scheme as a function of time.  The initial pulse, which is of
  compact support, completely leaves the domain shortly before $t=15$.
  The numerical errors left in the computational domain converge away
  at the correct rate.  The inset shows the $L_2$-norms of the errors
  at the different resolutions.}
\label{Fig:convtest}
\end{center}
\end{figure}

The coarse resolution is $200\times 200$ in each grid.  The fine
resolution is obtained by doubling the number of grid-points in each
direction.  The Cauchy slices extend from $r=2$, where an artificial
boundary is introduced to avoid the $r=0$ coordinate singularity, to
$r=10$, the location of the spherical interface boundary.  We use a
time step $\Delta t = \Delta u = 0.02$ for the coarse case and half it
for the fine case.  The test confirms that the scheme is indeed fourth
order convergent.  See figure \ref{Fig:convtest}.

\subsection{Tail decay}

\begin{figure}[ht]
\begin{center}
\includegraphics*[width=12cm]{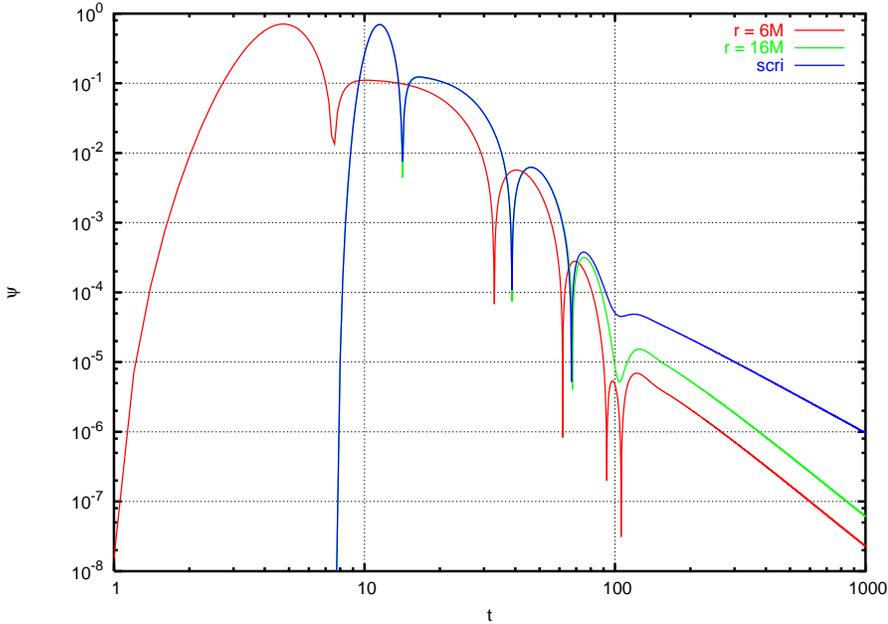}
\caption{Amplitude of the scalar field as measured by observers at
  fixed radial coordinate $r=6M$, $r=16M$ and at $\scri$.  Whereas the
  first observer is in the Cauchy region, the other two are in the
  characteristic region.  Note that the decay rate measured by the
  observer at infinity differs from that of the observers at a finite
  distance.  At $t=800M$ the ratio $d(\ln \Psi)/d(\ln t)$ is $-2.92$,
  for the $r=6M$ and $r=16M$ observers, and $-1.97$, for the observer
  at $\scri$.}
\label{Fig:decay}
\end{center}
\end{figure}

Having established that the scheme is fourth order accurate we want
to determine whether it can accurately produce the late time tail
decays \cite{Price,GPP,B}.  The field $\Psi$ is expected to decay as
$t^{-3}$ (along a time-like curve and assuming spherically symmetric
initial data), because of the scattering of the field off the
curvature of spacetime asymptotically far from the black hole.  We set
$M=1$, give initial data to $\Psi$ consisting of a spherically
symmetric pulse with compact support,
\[
\Psi(0,r,\theta) = \left\{ \begin{array}{ll}
\cos^6(r-4), & |r-4|\le \pi / 2,\\
0, & \mbox{otherwise},
\end{array}
\right.
\]
and set its time derivative equal to zero.  We use $r_i = \xi_i = 10$
and $\xi_{\infty} = 20$.  We monitor the value of $\Psi$ at three
different locations: $r=6$, $r=16$ ($\xi = 15$) and $\xi =
\xi_{\infty}$ (i.e., at $\scri$).  Using $400$ grid-points in the
radial direction in both grids and very few in the angular direction,
we obtain good agreement with the expected result.  See figure
\ref{Fig:decay}.  The errors in the angular direction are of round-off
order, i.e., $10^{-15}$ times smaller than the solution.

\section{Conclusion}
\label{Sec:conclusion}

To study gravitational radiation at future null infinity, where it is
unambiguously defined, numerical schemes have to evolve the fields up
to that surface.  An approach which would allow for this is
Cauchy-characteristic matching (CCM), whereby Cauchy and
characteristic evolutions are matched across a time-like world-tube.
This approach presents many advantages over other boundary conditions,
such as maximally dissipative or absorbing boundary conditions.  CCM
makes it possible, for example, to compute accurate waveform and
polarisation properties at infinity.  It is a highly computationally
efficient technique which can eliminate the undesirable reflections
and inaccuracies that other boundary conditions would introduce \cite{W}.

In this work we described a fourth order accurate algorithm for a
Cauchy-characteristic formulation of the wave equation written in
first order form.  By compactifying the radial coordinate in the
characteristic region we can evolve all the way to $\scri$.  We
carried out convergence tests showing that the scheme is fourth order
convergent and were able to reproduce the expected tail decay rate.
The interface was implemented using touching grids, which share a
common smooth boundary, and by communicating the relevant
characteristic/null variables.

The success for the scalar field case is clearly only a necessary
condition for the fully nonlinear case.  The nonlinear extension of
this work will require a careful choice of coordinate conditions and
attention to the propagation of constraints \cite{CLT,CPRST,SW,KW}.  It will be
crucial to establish how constraint violations propagate across the
domain, so that the interface treatment can be done consistently. In
three dimensions one would still require a smooth interface boundary
implemented through either overlapping grids \cite{CN2,Tho,CN1}, which
work well with simultaneous slices and can also allow for moving
boundaries, or touching grids \cite{LRT,SDDT,DDST}.

\section*{References}



\begin{thebibliography}{99}

\bibitem{P}
Pretorius F 2005 {\it  Phys.\ Rev.\  Lett.\ }{\bf 95} 121101

\bibitem{Goddard}
Baker J, Centrella J, Choi D, Koppitz M and van
   Meter J 2006 {\it Phys.\ Rev.\ Lett.} {\bf 96} 111102

\bibitem{Brownsville}
Campanelli M, Lousto C, Marronetti P and
   Zlochower Y 2006 {\it Phys.\ Rev.\ Lett.} {\bf 96} 111101

\bibitem{Lior}
Allen E, Buckmiller E, Burko L and Price R 2004 {\it Phys.\ Rev.\ D}
{\bf 70} 044038

\bibitem{DR}
Dafermos M and Rodnianski I 2004 {\it Preprint} gr-qc/0403034

\bibitem{Fra}
Frauendiener J 2004 {\it Living Rev.\ Relativity\ }{\bf 7} 1. URL (cited
on 7 April 2006) {\tt http://www.livingreviews.org/lrr-2004-1}


\bibitem{Kan} 
Kansagra A 2003 ``Coordinate compactifications and
hyperboloidal slices in numerical relativity'' LIGO/Caltech
undergraduate project, unpublished. For an online abstract, see
{\tt http://www.ligo.caltech.edu/LIGO\_web/students/ugprojects03.html}

\bibitem{M1} 
Misner C and Scheel M 2003 ``Wave propagation with
  hyperboloidal slicings'', talk given 12 June 2003 at KITP,
  unpublished. For an online version see
  {\tt http://online.kitp.ucsb.edu/online/gravity03/misner}

\bibitem{M2}
Misner C 2004 {\it Preprint} gr-qc/0409073

\bibitem{M3}
van Meter J, Fiske D and Misner C 2006 {\it Preprint} gr-qc/0603034

\bibitem{Sa}
Sachs R 1962 {\it Proc.\ R.\ Soc.\ London,
  Ser. A} {\bf 270} 103-126

\bibitem{BBM}
Bondi H, van der Burg M and Metzner A 1962 {\it Proc.\ R.\ Soc.\ London,
  Ser. A} {\bf 269} 21-52

\bibitem{W}
Winicour J 2001 {\it Living Rev.\  Relativity\ }{\bf 4} 3. URL (cited
on 7 April 2006) {\tt http://www.livingreviews.org/lrr-2001-3}

\bibitem{dI}
d'Inverno R 1992 {\em Approaches to Numerical Relativity} (Cambridge
University Press, Cambridge)

\bibitem{CdI}
Clarke C and d'Inverno R 1994 {\it Class.\ Quantum  Grav.} {\bf 11} (6) 1463-1448

\bibitem{CdIV}
Clarke C, d'Inverno R and Vickers J 1995 {\it Phys.\ Rev.\ }D {\bf 52}  (12) 6863-6867

\bibitem{DdIC}
Dubal M, d'Inverno R and Clarke C 1995 {\it Phys.\ Rev.\ }D {\bf 52}  (12) 6868-6881

\bibitem{dIV3}
d'Inverno R and Vickers J 1996 {\it Phys.\ Rev.\ }D {\bf 54}  (8) 4919-4928

\bibitem{dIV4}
d'Inverno R and Vickers J 1997 {\it Phys.\ Rev.\ }D {\bf 56}  (2) 772-784

\bibitem{DdIV}
Dubal M, d'Inverno R and Vickers J 1998 {\it Phys.\ Rev.\ }D {\bf 58}  044019

\bibitem{dIDS}
d'Inverno R, Dubal M and Sarkies E 2000 {\it Class. Quantum Grav.} {\bf 17}  (16) 3157-3170

\bibitem{SSV1}
Sperhake U, Sj\"odin K and Vickers J 2001 {\it Phys.\ Rev.\ }D {\bf 63} 024011

\bibitem{SSV2}
Sperhake U, Sj\"odin K and Vickers J 2001 {\it Phys.\  Rev.\ }D {\bf 63} 024012


\bibitem{BGHMPW1}
Bishop N, G\'omez R, Holvorcem P, Matzner R, Papadopoulos P and Winicour J 1996 {\it Phys.\ Rev.\ Lett.} {\bf 76} (23) 4303-4306

\bibitem{BGHMPW2}
Bishop N, G\'omez R, Holvorcem P, Matzner R, Papadopoulos P and Winicour J 1997 {\it J.\ Comput.\ Phys.} {\bf 136}  (1) 140-167

\bibitem{GLPW}
G\'omez R, Laguna P, Papadopoulos P and Winicour J 1996 {\it Phys.\ Rev.\ }D {\bf 54}  (8) 4719-4727


\bibitem{CGH}
Calabrese G, Gundlach C and Hilditch D 2006 {\it Class.\ Quantum
  Grav.} {\bf 23} 4829-4845 


\bibitem{SKW}
Szilagyi B, Kreiss H-O and Winicour J 2005 {\it Phys.\ Rev.\ D} {\bf
  71} 104035

\bibitem{MBSKW}
Motamed M, Babiuc M, Szilagyi B, Kreiss H-O and
   Winicour J 2006 {\it Phys.\ Rev.\ D} {\bf 73} 124008

\bibitem{C1}  
Calabrese G 2005 {\it Phys.\ Rev.\ D} {\bf 71} 027501(4)

\bibitem{CHH}
Calabrese G, Hinder I and Husa S 2005 {\it Preprint} gr-qc/0503056

\bibitem{CG} 
Calabrese G and Gundlach C 2006 {\it Class.\ Quantum Grav.} {\bf 23}
S343-S367 


\bibitem{CLNPRST}  
Calabrese G, Lehner L, Neilsen D, Pullin J, Reula O,
Sarbach O and Tiglio M 2003 {\it Class.\ Quantum Grav.\ }{\bf 20} L245-L252


\bibitem{St}
Strand B 1994 {\it J.\ Comput.\ Phys.\ }{\bf 110} 47

\bibitem{CLRST}  
Calabrese G, Lehner L, Reula O, Sarbach O, and Tiglio M 2004 {\it
  Class.\ Quantum Grav.\ }{\bf 21} 5735-5758 

\bibitem{Ols1} 
Olsson P 1995 {\it Math.\ Comp.\ } {\bf 64} 1035

\bibitem{Ols2} 
Olsson P 1995 {\it Math.\ Comp.\ } {\bf 64} S23

\bibitem{Ols3} 
Olsson P 1995 {\it Math.\ Comp.\ } {\bf 64} 1473

\bibitem{CGA}
Carpenter M, Gottlieb D and Abarbanel S 1994 {\it J.\ Comput.\
  Phys.} {\bf 111}

\bibitem{CNG}
Carpenter M, Nordstrom J and Gottlieb D 1999 {\it J.\ Comput.\
  Phys.} {\bf 148} 2 341

\bibitem {NC}
Nordstrom J and Carpenter M 2001 {\it J.\ Comput.\ Phys.} {\bf 173}

\bibitem{F1}
Frittelli S 2004 {\it J.\ Phys.\ A:  Math.\ Gen.\ }{\bf 37} 8639-8655

\bibitem{F2}
Frittelli S 2005 {\it J.\ Phys.\ A:   Math.\ Gen.\ }{\bf 38} 4209-4221

\bibitem{F3}
Frittelli S 2005 {\it  Phys.\ Rev.\ }D {\bf 71} 024021


\bibitem{GKO}
Gustafsson B, Kreiss H-O and Oliger J 1995
{\em Time dependent problems and difference methods}
(New York: Wiley)


\bibitem{CN2}  
Calabrese G and Neilsen D 2005 {\it Phys.\ Rev.\ D} {\bf 71} 124027(20)

\bibitem{Price}
Price R 1972 {\it Phys.\ Rev.\ D} {\bf 5} 2419

\bibitem{GPP}
Gundlach C, Price R and Pullin J 1994 {\it Phys.\ Rev.\ D} {\bf 49} 883

\bibitem{B}
Barack L 1999 {\it Phys.\ Rev.\ }D {\bf 59} 044017

\bibitem{CLT} 
Calabrese G, Lehner L and Tiglio M 2002 {\it Phys.\ Rev.\ D} {\bf 65} 104031(13)

\bibitem{CPRST} 
Calabrese G, Pullin J, Reula O, Sarbach O and Tiglio M 2003 {\it
  Commun.\ Math.\ Phys.\ }{\bf 240} 377-395 


\bibitem{SW}
Szilagyi B and Winicour J 2002 {\it Phys.\ Rev.\ D} {\bf 66} 064019

\bibitem{KW}
Kreiss H-O and Winicour J 2006 {\it Class.\ Quantum Grav.} {\bf 23}
S405-S420 

\bibitem{Tho}
Thornburg J 2004 {\it Class.\ Quantum Grav.\ }{\bf 21} 3665

\bibitem{CN1}  
Calabrese G and Neilsen D 2004 {\it Phys.\ Rev.\ D} {\bf 69} 044020(21)

\bibitem{LRT}
Lehner L, Reula O and Tiglio M 2005 {\it Class.\ Quantum Grav.\ }{\bf 22} 5283-5322


\bibitem{SDDT}
Schnetter E, Diener P, Dorband E and  Tiglio M 2006 {\it Class.\
  Quantum Grav.} {\bf 23} S553-S578 

\bibitem{DDST}
Diener P, Dorband E, Schnetter E and Tiglio M 2005 {\it Preprint} gr-qc/0512001


\end{thebibliography}
\end{document}